\documentclass{ws-procs9x6}

\begin{document}

\title{Heavy quark production and non-linear gluon evolution at the LHC
\footnote{Presented at the XIV International Workshop on Deep Inelastic Scattering, 
April 20-24, Tsukuba, Japan}}

\author{Krisztian Peters}

\address{School of Physics \& Astronomy, 
University of Manchester\\ Manchester, M13 9PL, UK\\ E-mail:
petersk@fnal.gov}

\maketitle

\abstracts{
We investigate the importance of unitarity corrections to parton
evolution in heavy flavor production at the LHC. The gluon distribution 
is determined with a fit to HERA data applying a unified
BFKL-DGLAP approach, in which the non-linear evolution is described by
the Balitsky-Kovchegov equation. First we estimate $b\bar b$
production at CDF and D0. Then, cross sections for heavy quark
production at various LHC experiments are estimated, tracing the
impact of the unitarity corrections.}

\section{Non-linear gluon evolution}

HERA measurements found a steep power-like growth of the gluon density 
with decreasing $x$ which would lead to a violation of unitarity at 
very small $x$ values. The steep growth has to be tamed by gluon 
rescattering, corresponding to non-linear effects in the gluon
density evolution. 

The standard framework to determine parton evolution is  
the collinear DGLAP formalism. It works rather well for inclusive 
quantities but, for more exclusive processes,
the $k_t$-factorization scheme is more appropriate because both the 
longitudinal and transverse components of the gluon momenta are considered.
In this framework, the process-independent quantity is the unintegrated 
gluon distribution, connected to the process-dependent hard 
matrix element via the $k_t$-factorization theorem.
Linear evolution of the unintegrated gluon distribution may be described
by one of the small~$x$ evolution equations using the 
$k_t$-factorization scheme, the BFKL~\cite{BFKL} and CCFM~\cite{CCFM} equations. 
These equations are based on resummation of large logarithmic pQCD
corrections, $\alpha_s^n \ln^m (1/x)$, and are equivalent at the 
leading logarithmic level.

The very small $x$ kinematic region is also the regime
where the growth of the gluon density must be tamed in order to preserve 
unitarity.
Recently, a successful description of unitarity corrections 
to DIS was derived within the color dipole formulation of QCD. 
This is the Balitsky-Kovchegov (BK) equation \cite{Bal,Kov} which 
describes the BFKL evolution of the gluon in a large target, including 
a non-linear term corresponding to gluon recombination 
at high density.

In our analysis, we determine the unintegrated gluon distribution
from the BK equation unified with the DGLAP equation following KMS
(Kwieci\'nski, Martin and 
Sta\'sto)~\cite{Kwiecinski:1997ee,Kimber:2001nm,Kutak:2003bd,Kutak:2004ym}.
We use the abbreviation KKMS (Kutak, Kwieci\'nski, Martin and 
Sta\'sto)~\cite{Kutak:2003bd,Kutak:2004ym} for the unified non-linear equation. 
The linear part of this equation is given by the BFKL kernel
with subleading $\ln(1/x)$ corrections, supplemented by the non-singular
parts of the DGLAP splitting functions. Thus resummation of both the
leading $\ln Q^2$ and $\ln (1/x)$ terms are achieved. 
The subleading terms in $\ln (1/x)$ are approximated by the so-called 
consistency constraint and the running coupling constant.
The non-linear part is taken directly from the BK equation, ensuring that 
the unitarity constraints are preserved. One expects that this framework
provides a more reliable description of the gluon evolution at 
extremely small~$x$, where $\ln (1/x) \gg 1$ and the unitarity corrections 
are important, than does DGLAP.

The size of the dense gluon system inside the proton
is assumed to be $R=2.8$~GeV$^{-1}$, in accord with the diffractive slope,
$B_d \simeq 4$~GeV$^{-2}$, of the elastic $J/\psi$ photoproduction cross
section at HERA. 
In this process, the impact parameter profile of the proton defines 
the $t$ dependence of the elastic cross section, 
$B_d \simeq R^2/2$, by Fourier transform. 

\section{Constraints from HERA and cross checks at the Tevatron}
%\label{sec:constr}

The initial distribution was obtained by fitting the
HERA $F_2$ measurements~\cite{Aid:1996au,Derrick:1995ef} 
using the Monte Carlo CASCADE \cite{Jung:2001hx,Jung:2000hk} for
evolution and convolution with the off-shell matrix elements.
The fits were repeated both with the standard KMS evolution without
the non-linear contribution and with extended KMS evolution including
the non-linear part. The predicted $F_2$ is equivalent for both linear
and non-linear evolution.

\begin{figure}[!h]
\begin{center}
\includegraphics[width=.4\textwidth]{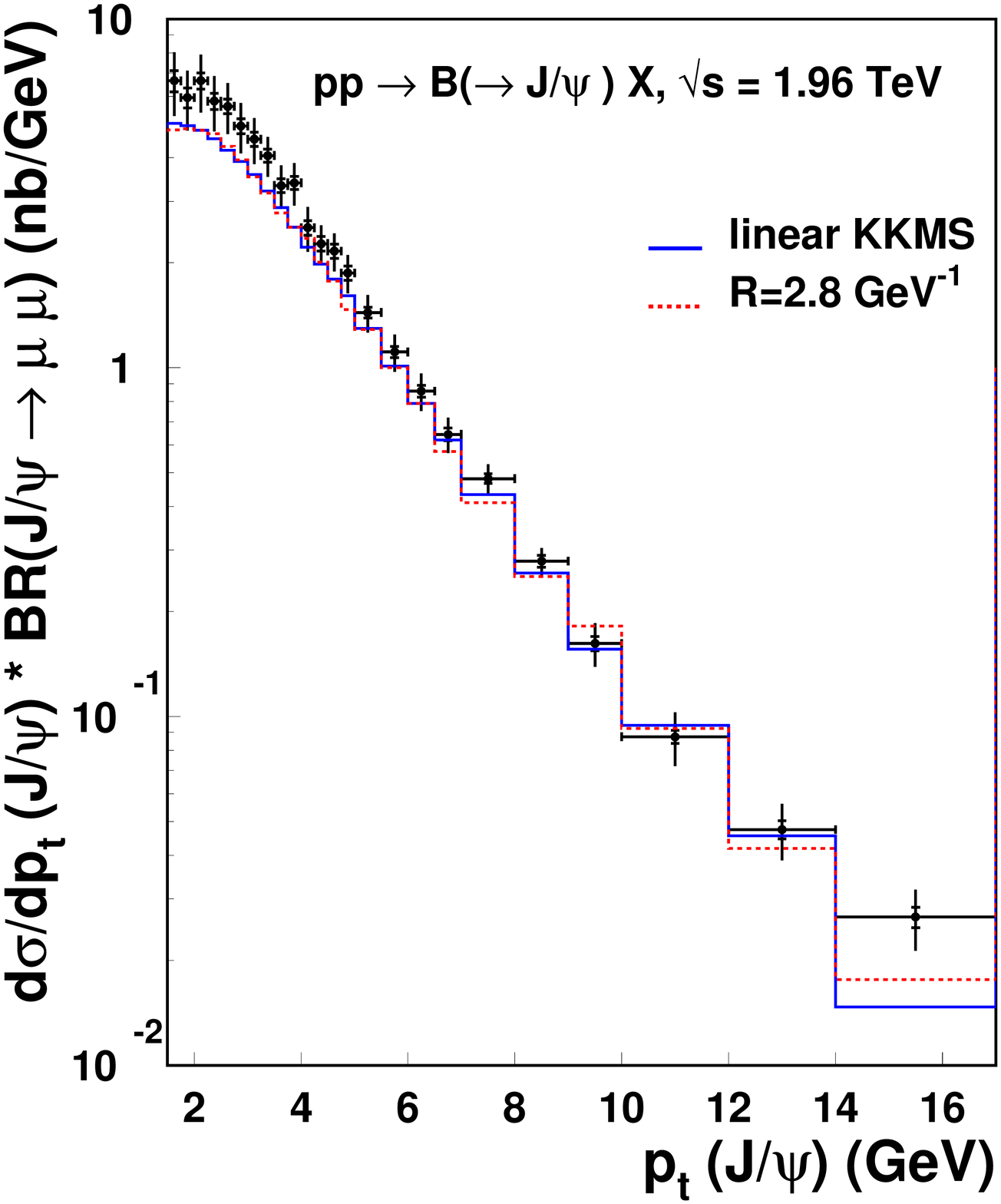}
\includegraphics[width=.5\textwidth]{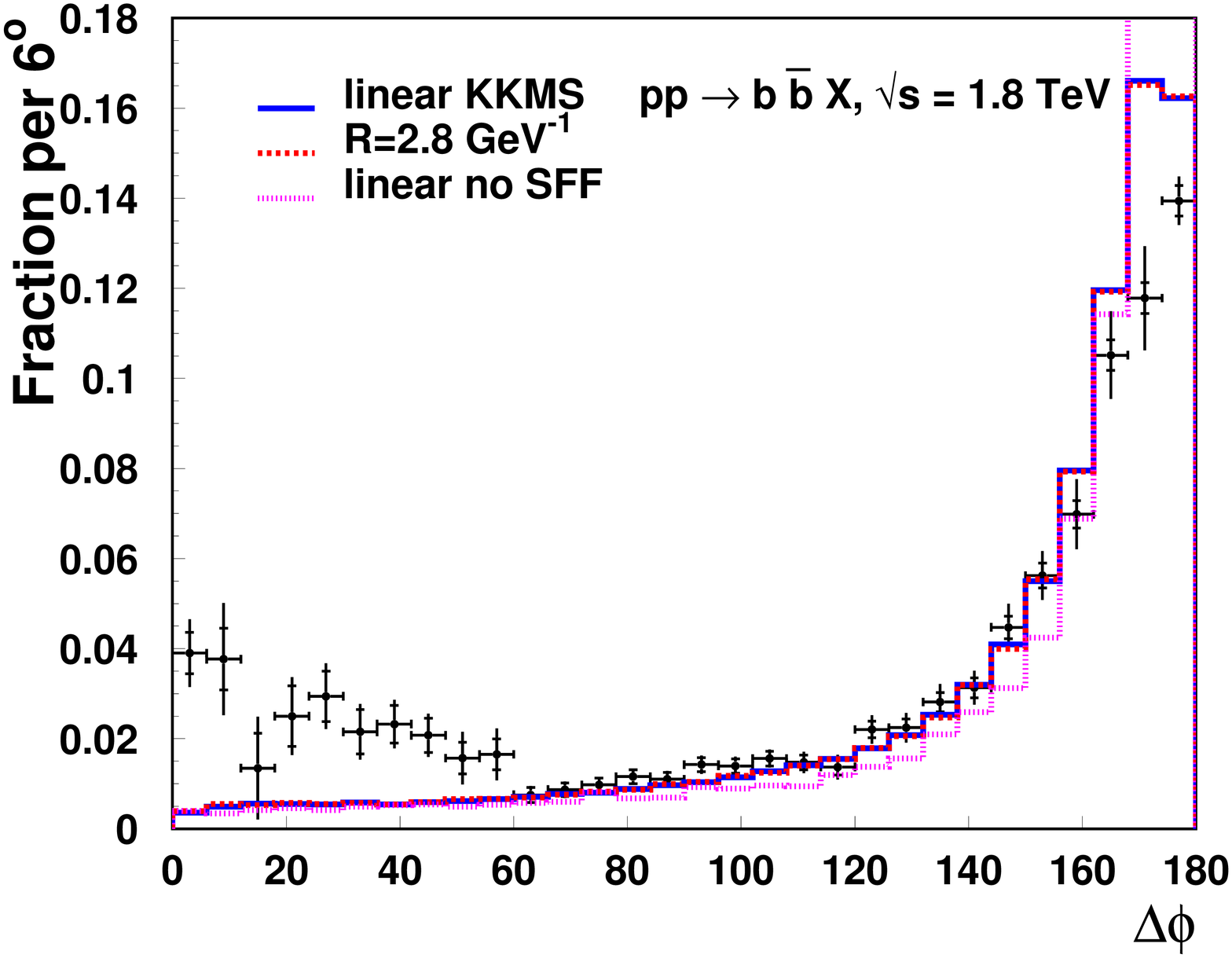}
\end{center}
\vskip -4cm 
\hskip 4cm $(a)$ \hskip 5cm $(b)$ 
\vskip 3cm
\caption{Bottom production, measured by CDF, is compared to predictions 
using CASCADE with linear and non-linear KKMS evolution. (a) The
$p_T$ distribution of $B$ meson decays to $J/\psi$.  (b) The azimuthal
angle, $\Delta \phi$, distribution of $b\bar b$ pair production
smeared by the experimental resolution.}
\label{fig:cdf}
\end{figure}

Next, this constrained gluon density was used to calculate the charm
structure function $F_2^c$ at HERA and $gg \rightarrow b\bar b$
production at the Tevatron as a cross-check of the fit and the
evolution formalism. We use $m_c=1.4$~GeV, $m_b=4.75$~GeV and a
renormalization scale in $\alpha_s$ of $Q^2 = 4m^2_q + p_T^2$. Sudakov
Form Factor is included in the calculation. 
The predicted cross section was then compared to both
H1 \cite{Adloff:2001zj}, Zeus
\cite{Breitweg:1999ad} and CDF \cite{Acosta:2004yw,Acosta:2004nj}, D0
\cite{Abbott:1999se} measurements respectively. The predictions agree
reasonably well with the data.
 
As an example of these cross-checks, in Fig.~\ref{fig:cdf}(a) the cross
section for $B$ decays to $J/\psi$ is shown as a function the $J/\psi$
$p_T$~\cite{Acosta:2004yw,Acosta:2004nj}. The KKMS gluon density fits
the data both in the linear and non-linear scenarios with a comparable
accuracy to the NLO collinear approach \cite{Cacciari:2003uh}.

In Fig.~\ref{fig:cdf}(b), the azimuthal angle distribution between the
$b$ and $\bar b$ quarks, $\Delta\phi$, is given. The $\Delta \phi$ and
$b \bar b$ $p_T$ distributions are correlated since $\Delta \phi <
180^\circ$ corresponds to higher pair $p_T$.  Since the
$k_T$-factorization formula allows the incoming gluons to have sizable
transverse momenta, the calculated $\Delta \phi$ distribution agrees
very well with the data for $\Delta \phi > 60^\circ$ with only
smearing due to the experimental resolution. It is interesting to note
that the inclusion of the Sudakov From Factor improves the description
significantly. For a comparison, we also plotted the result as obtained
without the Sudakov Form Factor (dotted line). The enhancement of the
data relative to the calculations at low $\Delta \phi$ requires
further study.

\section{Heavy quark production at the LHC}

\begin{figure}
\begin{center}
\hspace{-.6cm}\includegraphics[width=.34\textwidth]{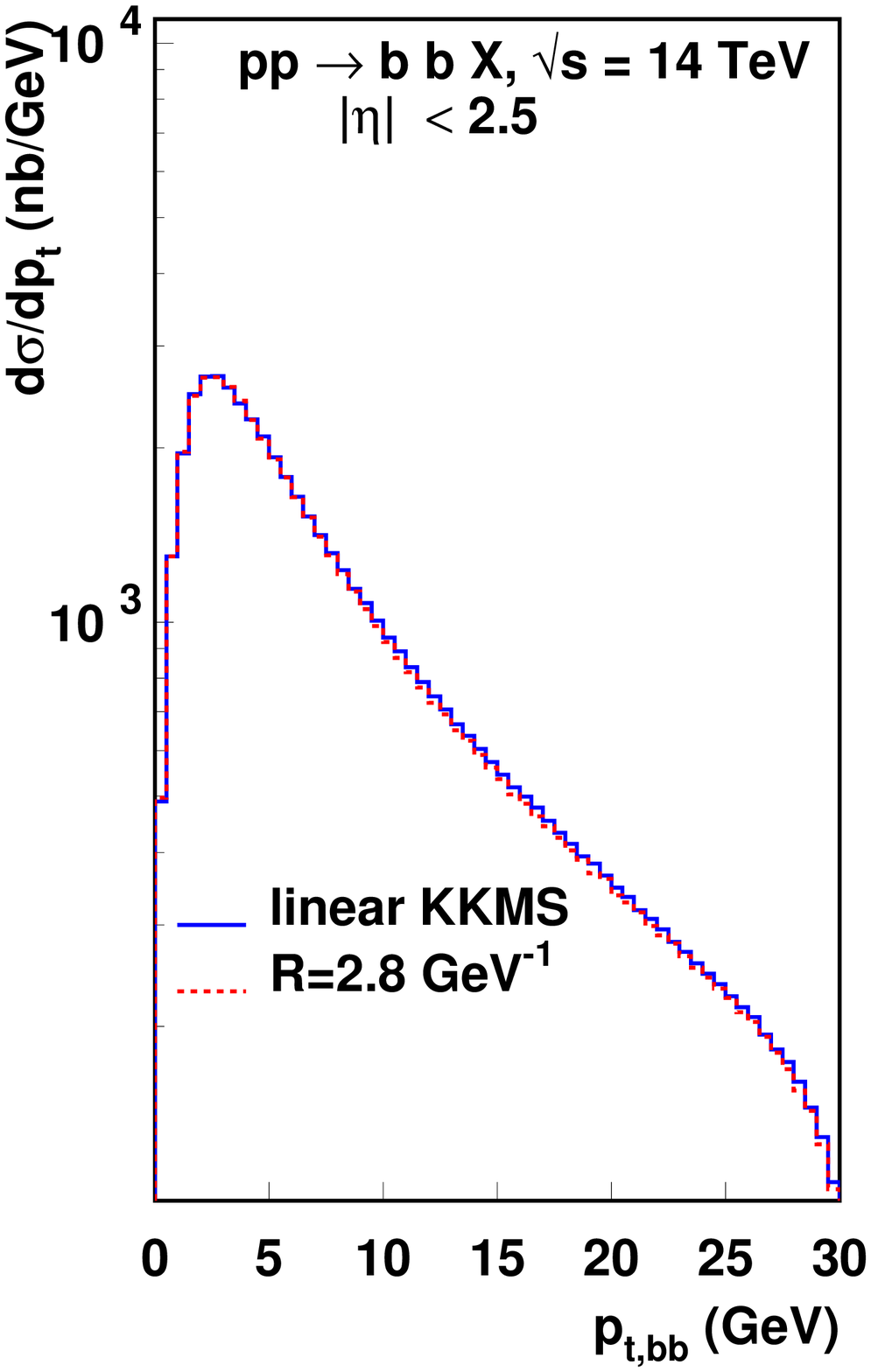}
\hspace{-.6cm}\includegraphics[width=.34\textwidth]{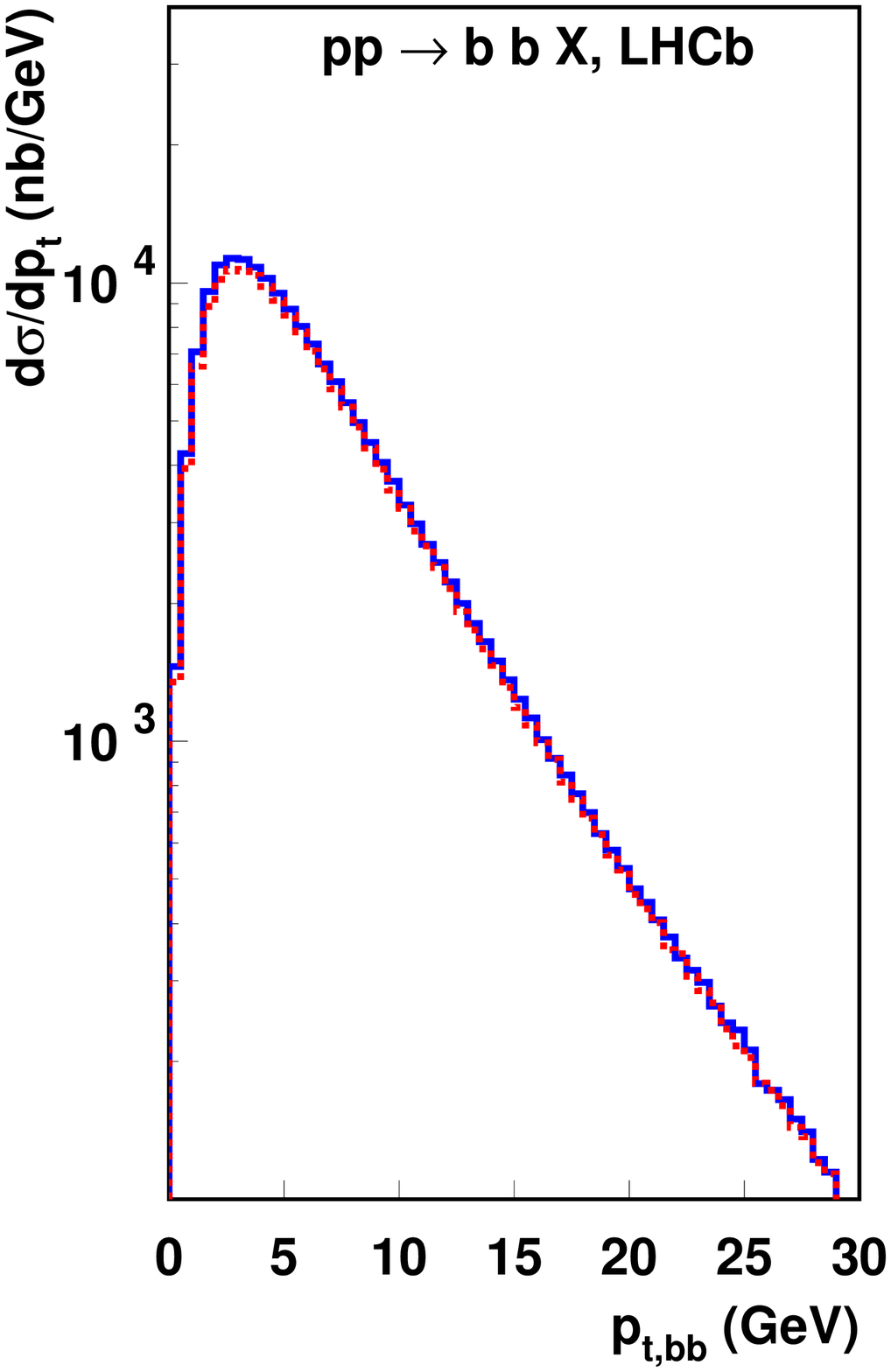}
\hspace{-.6cm}\includegraphics[width=.34\textwidth]{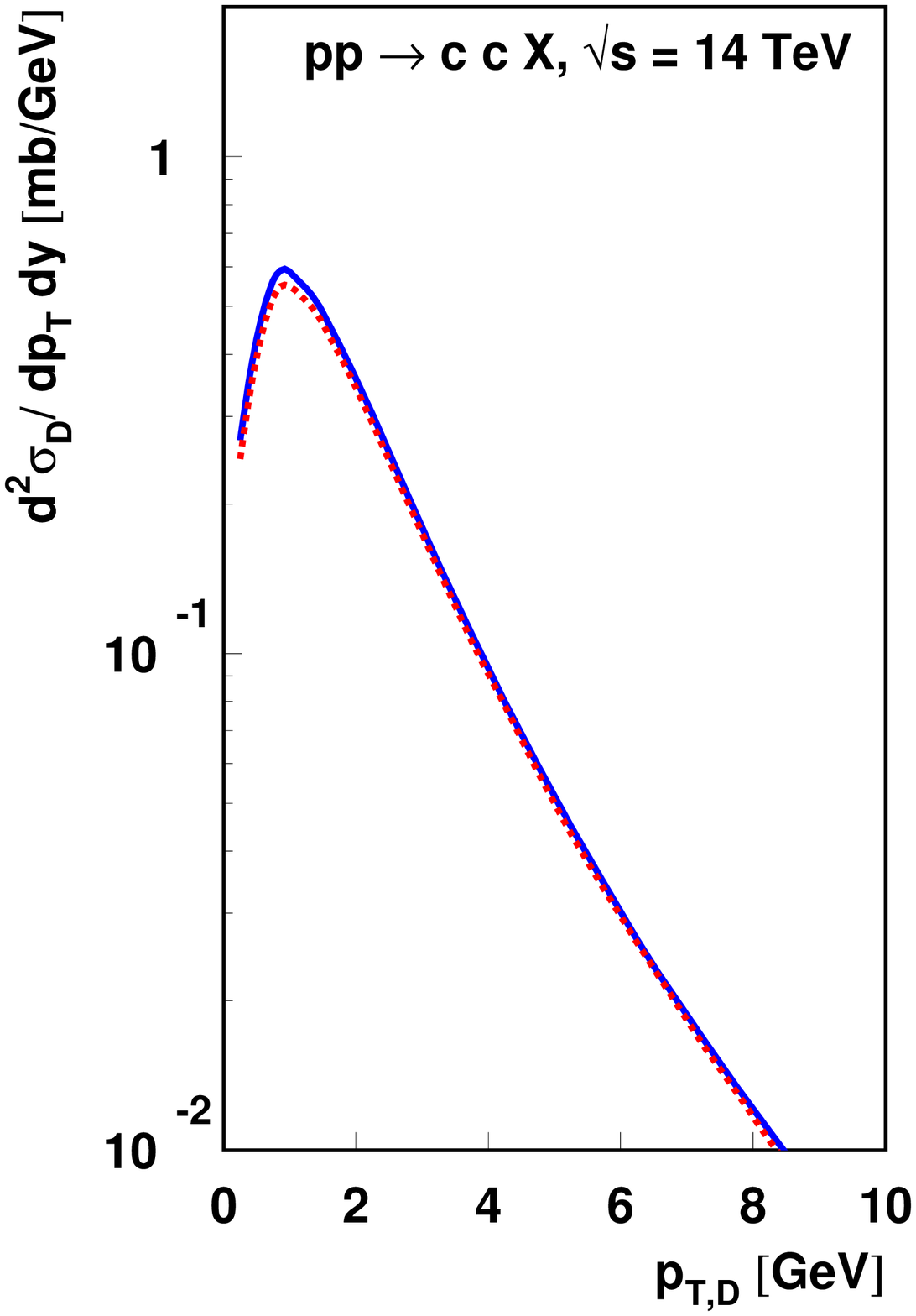}
\end{center}
\vskip -3.5cm 
\hskip 2.7cm $(a)$ \hskip 2.85cm $(b)$ \hskip 2.8cm $(c)$
\vskip 2.4cm
\caption{(a) and (b) show $b \overline b$ production as a function of pair
$p_T$ in the ATLAS/CMS (a) and the LHCb acceptance (b).
The $D^0$ meson $p_T$ distribution in the ALICE acceptance is shown in (c).}
\label{fig:lhc}
\end{figure}

Since the Tevatron measurements are well described using the 
unintegrated parton densities constrained by HERA and
convoluted with the off-shell matrix elements, the same approach may be used 
for  heavy quark production at the LHC at {\it e.g.} $\sqrt s =14$ TeV. 
As discussed previously, heavy quark production at this 
energy is already in the region where saturation effects may be relevant. 

We computed heavy quark cross sections for various kinematical regions
of the LHC.  In Fig.~\ref{fig:lhc}(a), the $b\bar b$ production cross
section is computed within the ATLAS and CMS acceptance ($p_T > 10$
GeV and $|\eta | < 2.5 $ for both the $b$ and $\bar b$ quarks). In
Fig.~\ref{fig:lhc}(b), the same cross section is computed within the
LHCb acceptance where the $b$ quark $p_T$ can be measured to 2 GeV for
$1.9 <\eta < 4.9$. Similarly, we investigated $c\bar c$ production at
ALICE, Fig.~\ref{fig:lhc}(c).  In ALICE, it will be possible to
measure the $D^0$ down to $p_T \sim 0.5$ GeV in $|\eta | < 0.9 $. In
the computations the same quark masses and scale was used as
described in the previous section.

In all three cases the results of the linear evolution (solid line)
and the results of the non-linear evolution (dashed line) are very
similar.  There is no significant effect observable for non-linear
evolution due to gluon saturation. For $c\bar c$ production at ALICE
saturation effects have been predicted \cite{alicepap} within the GLR
approach \cite{glr} in the collinear limit (even with a larger
saturation radius). This result \cite{alicepap} could not be confirmed
with our calculations.

The presented results, Fig.~\ref{fig:lhc}, suggests that linear gluon
evolution and $k_T$-factorization can safely be applied in the
discussed kinematical regions of the LHC.

\section*{Acknowledgments}
The presented work was done in collaboration with H.~Jung, K.~Kutak and L.~Motyka.


\begin{thebibliography}{0}

\bibitem{BFKL}   L. N. Lipatov, {\it Sov. J. Nucl. Phys.}
                  {\bf 23} (1976) 338;\\
                  E. A. Kuraev, L. N. Lipatov and  V. S. Fadin,
                  {\it Sov. Phys. JETP} {\bf 45} (1977) 199;\\
                  I. I. Balitsky and  L. N. Lipatov,
                  {\it Sov. J. Nucl. Phys.} {\bf 28} (1978) 338

\bibitem{CCFM}
M.~Ciafaloni, {\it Nucl. Phys.} {\bf B 296} (1988) 49;
S. Catani, F. Fiorani and G.~Marchesini,
\newblock Phys. Lett.{} {\bf B 234} (1990) 339;
S.~Catani, F. Fiorani and G.~Marchesini,
\newblock Nucl. Phys.{} {\bf B 336} (1990) 18;
G.~Marchesini,
\newblock Nucl. Phys.{} {\bf B 445} (1995) 49

\bibitem{Bal}  I. I. Balitsky,  {\it Nucl. Phys.} {\bf  B463} (1996) 99;
{\it Phys. Rev. Lett.} {\bf 81} (1998) 2024;
              { \it Phys. Rev.} {\bf D60} (1999)  014020;
{\it  Phys. Lett.} {\bf B518} (2001) 235

\bibitem{Kov}  Yu. V. Kovchegov, {\it Phys. Rev.} {\bf D60} (1999) 034008

\bibitem{Kwiecinski:1997ee}
J. Kwiecinski, A.~D. Martin and A.~M. Stasto,
\newblock Phys. Rev.{} {\bf D 56}, (1997) 3991

\bibitem{Kimber:2001nm}
M.~A. Kimber, J. Kwiecinski and A.~D. Martin,
\newblock Phys. Lett.{} {\bf B 508} (2001) 58

\bibitem{Kutak:2003bd}
K. Kutak and J. Kwiecinski,
\newblock Eur. Phys. J.{} {\bf C 29} (2003) 521

\bibitem{Kutak:2004ym}
K. Kutak and A.~M. Stasto,
\newblock Eur. Phys. J.{} {\bf C 41} (2005) 343

%\cite{Aid:1996au}
\bibitem{Aid:1996au}
  S.~Aid {\it et al.}  [H1 Collaboration],
  %``A Measurement and QCD Analysis of the Proton Structure Function
  %$F_2(x,Q~2)$ at HERA,''
  Nucl.\ Phys.\ B {\bf 470} (1996) 3
  %[arXiv:hep-ex/9603004].
  %%CITATION = HEP-EX 9603004;%%

%\cite{Derrick:1995ef}
\bibitem{Derrick:1995ef}
  M.~Derrick {\it et al.}  [ZEUS Collaboration],
  %``Measurement of the Proton Structure Function ${F_2}$ at low ${x}$ and low
  %${Q~2}$ at HERA,''
  Z.\ Phys.\ C {\bf 69} (1996) 607
  %[arXiv:hep-ex/9510009].
  %%CITATION = HEP-EX 9510009;%%

%\cite{Jung:2001hx}
\bibitem{Jung:2001hx}
  H.~Jung,
  %``The CCFM Monte Carlo generator CASCADE,''
  Comput.\ Phys.\ Commun.\  {\bf 143} (2002) 100
  %[arXiv:hep-ph/0109102].
  %%CITATION = HEP-PH 0109102;%%  Phys.\ Lett.\ B {\bf 528}, 199 (2002)
  %[arXiv:hep-ex/0108039].
  %%CITATION = HEP-EX 0108039;%%

%\cite{Jung:2000hk}
\bibitem{Jung:2000hk}
  H.~Jung and G.~P.~Salam,
  %``Hadronic final state predictions from CCFM: The hadron-level Monte  Carlo
  %generator CASCADE,''
  Eur.\ Phys.\ J.\ C {\bf 19} (2001) 351
  %[arXiv:hep-ph/0012143].
  %%CITATION = HEP-PH 0012143;%%

\bibitem{Adloff:2001zj}
  C.~Adloff {\it et al.}  [H1 Collaboration],
  %``Measurement of D*+- meson production and F2(c) in deep inelastic
  %scattering at HERA,''
  Phys.\ Lett.\ B {\bf 528} (2002) 199
  %[arXiv:hep-ex/0108039].
  %%CITATION = HEP-EX 0108039;%%


\bibitem{Breitweg:1999ad}
  J.~Breitweg {\it et al.}  [ZEUS Collaboration],
  %``Measurement of D*+- production and the charm contribution to F2 in  deep
  %inelastic scattering at HERA,''
  Eur.\ Phys.\ J.\ C {\bf 12} (2000) 35
  %[arXiv:hep-ex/9908012].
  %%CITATION = HEP-EX 9908012;%%

%\cite{Acosta:2004yw}
\bibitem{Acosta:2004yw}
  D.~Acosta {\it et al.}  [CDF Collaboration],
  %``Measurement of the J/psi meson and b-hadron production cross sections  in p
  %anti-p collisions at s**(1/2) = 1960-GeV,''
  Phys.\ Rev.\ D {\bf 71} (2005) 032001
  %[arXiv:hep-ex/0412071].
  %%CITATION = HEP-EX 0412071;%%

%\cite{Acosta:2004nj}
\bibitem{Acosta:2004nj}
  D.~Acosta {\it et al.}  [CDF Collaboration],
  %``Measurements of bottom anti-bottom azimuthal production correlations in
  %proton antiproton collisions at s**(1/2) = 1.8-TeV,''
  Phys.\ Rev.\ D {\bf 71} (2005)  092001
  %[arXiv:hep-ex/0412006].
  %%CITATION = HEP-EX 0412006;%%


%\cite{Abbott:1999se}
\bibitem{Abbott:1999se}
  B.~Abbott {\it et al.}  [D0 Collaboration],
  %``The b anti-b production cross section and angular correlations in p  anti-p
  %collisions at s**(1/2) = 1.8-TeV,''
  Phys.\ Lett.\ B {\bf 487} (2000) 264
  %[arXiv:hep-ex/9905024].
  %%CITATION = HEP-EX 9905024;%%

\bibitem{Cacciari:2003uh}
M. Cacciari, S. Frixione, M.~L. Mangano, P. Nason and G. Ridolfi,
\newblock JHEP{} {\bf 07} (2004) 033


\bibitem{glr} L.~V.~Gribov, E.~M.~Levin and M.~G.~Ryskin,
  %``Semihard Processes In QCD,''
  {\it Phys.\ Rept.\ } {\bf 100} (1983) 1

\bibitem{alicepap}
A. Dainese, R. Vogt, M. Bondila, K.~J. Eskola and V.~J. Kolhinen,
\newblock J. Phys.{} {\bf G 30} (2004) 1787


\end{thebibliography}
\end{document}